\documentclass[notitlepage,twocolumn,showpacs,10pt]{revtex4-1}
\usepackage[utf8]{inputenc}
\usepackage{amsmath}
\usepackage{amsfonts}
\usepackage{amssymb}
\usepackage{times,fullpage}
\usepackage{comment}
\usepackage{array,bm}
\usepackage{graphicx,tikz}

\begin{document}

\newcommand{\nwc}{\newcommand}
\nwc{\vs}{\vspace}
\nwc{\hs}{\hspace}
\nwc{\la}{\langle}
\nwc{\ra}{\rangle}
\nwc{\nn}{\nonumber}
\nwc{\Ra}{\Rightarrow}
\nwc{\wt}{\widetilde}
\nwc{\lw}{\linewidth}
\nwc{\ft}{\frametitle}
\nwc{\ben}{\begin{enumerate}}
\nwc{\een}{\end{enumerate}}
\nwc{\bit}{\begin{itemize}}
\nwc{\eit}{\end{itemize}}
\nwc{\dg}{\dagger}
\nwc{\mA}{\mathcal A}
\nwc{\mD}{\mathcal D}
\nwc{\mB}{\mathcal B}

\nwc{\Tr}[1]{\underset{#1}{\mbox{Tr}}~}
\nwc{\pd}[2]{\frac{\partial #1}{\partial #2}}
\nwc{\ppd}[2]{\frac{\partial^2 #1}{\partial #2^2}}
\nwc{\fd}[2]{\frac{\delta #1}{\delta #2}}
\nwc{\pr}[2]{K(i_{#1},\alpha_{#1}|i_{#2},\alpha_{#2})}
\nwc{\av}[1]{\left< #1\right>}
\nwc{\cb}[1]{\textcolor{blue}{#1}}

\nwc{\zprl}[3]{Phys. Rev. Lett. ~{\bf #1},~#2~(#3)}
\nwc{\zpre}[3]{Phys. Rev. E ~{\bf #1},~#2~(#3)}
\nwc{\zpra}[3]{Phys. Rev. A ~{\bf #1},~#2~(#3)}
\nwc{\zjsm}[3]{J. Stat. Mech. ~{\bf #1},~#2~(#3)}
\nwc{\zepjb}[3]{Eur. Phys. J. B ~{\bf #1},~#2~(#3)}
\nwc{\zrmp}[3]{Rev. Mod. Phys. ~{\bf #1},~#2~(#3)}
\nwc{\zepl}[3]{Europhys. Lett. ~{\bf #1},~#2~(#3)}
\nwc{\zjsp}[3]{J. Stat. Phys. ~{\bf #1},~#2~(#3)}
\nwc{\zptps}[3]{Prog. Theor. Phys. Suppl. ~{\bf #1},~#2~(#3)}
\nwc{\zpt}[3]{Physics Today ~{\bf #1},~#2~(#3)}
\nwc{\zap}[3]{Adv. Phys. ~{\bf #1},~#2~(#3)}
\nwc{\zjpcm}[3]{J. Phys. Condens. Matter ~{\bf #1},~#2~(#3)}
\nwc{\zjpa}[3]{J. Phys. A ~{\bf #1},~#2~(#3)}
\nwc{\zpjp}[3]{Pramana J. Phys. ~{\bf #1},~#2~(#3)}

\title{Extended Fluctuation Theorems for Repeated Measurements and Feedback within Hamiltonian Framework}
\author{Sourabh Lahiri$^1$} 
\email{sourabhlahiri@gmail.com}
 \author{A. M. Jayannavar$^2$}
\email{jayan@iopb.res.in}
\affiliation{$^1$ Laboratoire de Physico-Chimie Th\'eorique, ESPCI, 10 rue Vauquelin, F-75231 Paris, France\\
$^2$Institute of Physics, Sachivalaya Marg, Bhubaneswar 751005, India}

\begin{abstract}
We derive the extended fluctuation theorems in presence of multiple measurements and feedback, when the system is governed by Hamiltonian dynamics. We use only the forward phase space trajectories in the derivation. However, to obtain an expression for the efficacy parameter, we must necessarily use the notion of reverse trajectory. Our results show that the correction term appearing in the exponent of the extended fluctuation theorems are non-unique, whereas the physical meaning of the efficacy parameter is unique.
\end{abstract}

\maketitle
\section{Introduction}

The field of nonequilibrium statistical mechanics, although vastly studied, has few exact laws that remain valid when the system of interest is far from thermodynamic equilibrium (i.e., the regime where the linear response theory breaks down). The so-called \emph{fluctuation theorems} are one of those rare equalities that have qualified to be in this list \cite{sei12_rpp,eva94_pre,jar97_prl,jar97_pre,cro98_jsp,cro99_pre,eva02_ap,sei05_prl,sei08_epjb,kur07_jsm,jar10_arcmp,lah09_pre,lah12a_jpa,lah14_epjb,lah14a_epjb}.
 One of them is the celebrated Jarzynski equality (JE) \cite{jar97_prl,jar97_pre}, which is relevant when the system is initially (at time $t=0$) at thermal equilibrium with a medium at temperature $\beta^{-1}$, and has thereby been perturbed by an external time-dependent perturbation $\lambda(t)$ for a time duration $\tau$. We take the average of the quantity $e^{-\beta W}$ (where $W$ is the work done \emph{on} the system) over a large number of experimental realizations of the same process. If $\Delta F$ is the net change in Hemholtz free energy, which is in turn equal to the work done during an isothermal reversible process, then the JE states that 
\begin{align}
\av{e^{-\beta (W-\Delta F)}}=1.
\label{eq:JE}
\end{align}
This, in conjunction with the Jensen's Inequality leads to the statement of the Second Law: $\av{W}\ge \Delta F$. In other words, the {\it average} work done on the system for any process cannot be less than that done during a reversible process. 

In recent literature, it has been shown that this equation as well as the corresponding Second Law inequality undergoes a change when the external perturbation is governed by a feedback mechanism \cite{sag10_prl,hor10_pre,pon10_pre,sag12_prl,lah12_jpa,lah12_pramana}. These results can be readily derived by starting with the ratio between the forward and time-reversed phase space paths in the forward process (with protocol $\lambda(t)$) and the reverse process (with protocol $\lambda(\tau-t)$), respectively. {Stated differently, the feedback control acts as a Maxwell's Demon that increases the free energy of the system \cite{sag11_jp}.}
However, the definition of the reverse process in presence of feedback is not unique. Depending on whether we choose to apply feedback during the reverse process, the  extended JE takes up different correction terms \cite{lah13_phyA}. 
But the reverse process is only a mathematical tool to prove the integral fluctuation relations, which in turn are averages over the forward paths only. This means that in an experimental verification of the theorems, we need not generate the reverse process at all. The question that one might ask is: can the theorems be proven without taking help of the reverse paths, using the expressions for the forward trajectories only? We show below that this can indeed be done for a system whose dynamics is governed by the Hamilton's equations of motion. The Hamiltonian approach has been used extensively and successfully in deriving the Jarzynski Equality, the Crooks Fluctuation Theorem, measure of dissipation and arrow of time. Hamiltonian derivation has also been used in \cite{sag11_jp} for single measurement and feedback. We extend this Hamiltonian derivation to the case of multiple measurement and feedback, using only the forward trajectories. However, for calculating the efficacy parameter, the extended FTs require the notion of  time-reversed trajectories, as detailed later.

\section{Jarzynski Equality for a Hamiltonian system with feedback}

We consider a system that is described by the Hamiltonian $H(t)\equiv H(\bm q,\bm p;t)$. A microstate of the system is given by a point $\bm x\equiv (\bm q,\bm p)$ in phase space, {whose evolution follows Hamiltonian dynamics:
\begin{align}
  \dot{\bm q}(t) &= \partial_{\bm p}H(t),\nn\\
  \dot{\bm p}(t) &= -\partial_{\bm q}H(t).
\end{align}
In the above equations, the overhead dot appearing on the LHS represents total derivative with respect to time, while the RHS consists of partial derivatives with respect to momenta and coordinates, respectively. Under the Hamiltonian dynamics, any phase point $\bm x_t$ can be written in terms of the initial phase point as $\bm x_t=M_t(\bm x_0)$, which is a one-to-one mapping \cite{sag11_jp}. Similarly, the inverse mapping can also be defined as $\bm x_0=M_t^{-1}(\bm x_t)$.}

 The system was initially in a state of thermal equilibrium, so that the initial state of the system follows the Boltzmann distribution
\begin{align}
p_0(\bm q_0, \bm p_0) = \frac{e^{-\beta H(0)}}{Z(0)}.
\end{align}
This condition can be relaxed for arbitrary initial distribution which will be discussed later.

We now subject this system to the following process \cite{lah12_jpa}: we measure the state of the system at time $t=0$ and obtain an outcome $\bm y_0$ with the error probability $p(\bm y_0|\bm x_0)$. Here, $\bm x_0$ is the actual initial state of the system, which is in general different from the measured outcome $\bm y_0$ due to measurement errors. Next, we apply an external protocol $\lambda_{\bm y_0}(t)$ from time $t=0$ up to time $t=t_1$, when a second measurement is performed. Once again, depending on the outcome $\bm y_1$ that is obtained with error probability $p(\bm y_1|\bm x_1)$, the functional form of the protocol is changed to $\lambda_{\hat{\bm y}_1}(t)$. Here, the subscript $\hat{\bm y}_1$ implies that the new protocol in general depends on the measurement history $\{\bm y_0,\bm y_1\}$. This process is continued till time $t_N$ when the last measurement is performed. The final value of the protocol is $\lambda_{\hat{\bm y}_N}(\tau)$. 


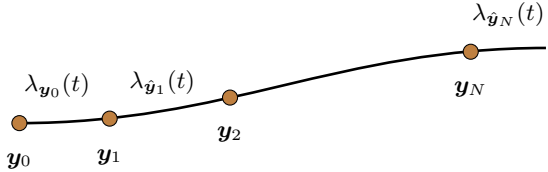
\begin{figure}[!h]
\centering
\begin{tikzpicture}
\draw [line width=1] (0,0) to [out=0,in=180] (7,1);
\draw [fill=brown] (0,0) circle (0.1cm);
\draw (0,-0.5) node {$\bm y_0$};
\draw (0.5,0.5) node {$\lambda_{\bm y_0}(t)$};

\draw [fill=brown] (1.2,0.06) circle (0.1cm);
\draw (1.2,0.05-0.5) node {$\bm y_1$};
\draw (1.2+0.7,0.05+0.5) node {$\lambda_{\hat{\bm y}_1}(t)$};

\draw [fill=brown] (2.8,0.34) circle (0.1cm);
\draw (2.8,0.35-0.5) node {$\bm y_2$};

\draw [fill=brown] (6,0.95) circle (0.1cm);
\draw (6,0.95-0.5) node {$\bm y_N$};
\draw (6+0.5,0.95+0.5) node {$\lambda_{\hat{\bm y}_N}(t)$};
\end{tikzpicture}
\caption{Protocol $\lambda(t)$ for a given set of measurements $\{\bm y_i\}$.}
\end{figure}

We will see that the JE gets modified in the presence of feedback. The modified relation is given by
\begin{align}
\av{e^{-\beta (W-\Delta F)-I}}=1.
\label{eq:MJE}
\end{align}
Here, $I$ is the {\it mutual information} between the actual variables $\{\bm x_i\}$ and the measured variables $\{\bm y_i\}$, defined through the expression \cite{hor10_pre,lah12_jpa}
\begin{align}
I(\{\bm x_i\},\{\bm y_i\}) = \ln\frac{p(\bm y_0|\bm x_0)p(\bm y_1|\bm x_1)\cdots p(\bm y_N|\bm x_N)}{P(\{\bm y_i\})}.
\end{align}
{Here, $P(\{y_i\})$ is the probability of a set of measurement outcomes $\{y_i\}$.}
The average is over all paths, the probability of each path being
\begin{align}
&p(\bm x_0) P_{\bm y_0}[\bm x_0 \to \bm x_1] p(\bm y_1|\bm x_1) P_{\hat{\bm y}_1}[\bm x_1 \to \bm x_2]\nn\\
&~~~~~ \times \cdots \times p(\bm y_N|\bm x_N) P_{\hat{\bm y}_N}[\bm x_N \to \bm x_\tau]\nn\\
&= p(\bm x_0)p(\bm y_0|\bm x_0)p(\bm y_1|\bm x_1)\cdots p(\bm y_N|\bm x_N)P_{\{y_i\}}(\{\bm x_i\}|\bm x_0).
\end{align}
This probability includes variables $\{\bm x_i\}$ as well as measured values $\{\bm y_i\}$. We assume that the different measurements are independent.
 $P_{\hat{\bm y}_k}[\bm x_k \to \bm x_{k+1}]$  is the probability for going from $\bm x_k$ to $\bm x_{k+1}$ under the protocol $\lambda_{\hat{\bm y}_k}(t)$. $P_{\{y_i\}}(\{\bm x_i\}|\bm x_0)$ is the probabability of the path $\{\bm x_i\}$ under the given protocol $\{\bm y_i\}$, with the given initial point $\bm x_0$.

To derive Eq.  \eqref{eq:MJE}, we first note that in a Hamiltonian dynamics, the total work done is equal to the net change in the internal energy: $H(\bm x_\tau,\tau)-H(\bm x_0,0)$. This follows from the fact that the net work done is equal to the sum of the works done in the intervals between any two measurements: $W=\sum_iW_i=\sum_i[H(\bm x_i,t_i)-H(\bm x_{i-1},t_{i-1})]$. For simplicity, from now on we will use the notation $H(t)\equiv H(\bm x(t),t)$.
We begin with the definition of the average appearing on the left hand side of Eq.  \eqref{eq:MJE}
\begin{align}
&\av{e^{-\beta (W-\Delta F)-I}} \nn\\
&= \int \{\bm{dx}_i\}\{ \bm{dy}_i\} p(\bm x_0)P_{\{\bm y_i\}}(\{\bm x_i\}|\bm x_0)\nn\\ 
&\hspace{0.5cm}\times p(\bm y_0|\bm x_0)p(\bm y_1|\bm x_1)\cdots p(\bm y_N|\bm x_N)~e^{-\beta (W-\Delta F)-I}.
\end{align}
Using the equilibrium initial distribution, it becomes
\begin{align}
& \int \{\bm{dx}_i\}\{ \bm{dy}_i\} \frac{e^{-\beta H(0)}}{Z(0)}P_{\{\bm y_i\}}(\{\bm x_i\}|\bm x_0)\nn\\ 
& \hspace{0.5cm}\times p(\bm y_0|\bm x_0)p(\bm y_1|\bm x_1)\cdots p(\bm y_N|\bm x_N) e^{-\beta (H(\tau)-H(0))}\nn\\
& \hspace{0.5cm}\times \frac{Z(0)}{Z(\tau)}
\times  \frac{P(\{\bm y_i\})}{p(\bm y_0|\bm x_0)p(\bm y_1|\bm x_1)\cdots p(\bm y_N|\bm x_N)}.
\end{align}
We have used the definition of $I$ and of free energy change in terms of the partition function: $\Delta F = \ln[Z(0)/Z(\tau)]$. 
Now, we make use of the fact that in Hamiltonian dynamics, $P_{\{\bm y_i\}}(\{\bm x_i\}|\bm x_0) = \delta(\bm x_1-M^{\hat{\bm y}_1}_{t_1}(\bm x_0))\delta(\bm x_2-M^{\hat{\bm y}_2}_{t_2}(\bm x_0))\cdots$. This means that once the initial point is given, all other phase points on the trajectory are determined and the trajectory is unique. This allows us to write
\begin{align}
\av{e^{-\beta (W-\Delta F)-I}}&= \int \bm{dx}_0\{ \bm{dy}_i\} \frac{e^{-\beta H(\tau)}}{Z(\tau)}P(\{\bm y_i\})\nn\\
&= \int \bm{dx}_\tau\{ \bm{dy}_i\} \frac{e^{-\beta H(\tau)}}{Z(\tau)}P(\{\bm y_i\})\nn\\
&= \frac{1}{Z(\tau)}\int \bm{dx}_\tau e^{-\beta H(\tau)} = 1.
\label{eq:proof}
\end{align}
In the last line, we have used the fact that the Jacobian for transformation of $\bm x_0$ to $\bm x_\tau$ is unity: $\bm{dx}_0=\bm{dx}_\tau$.
Thus, we have derived eq. \eqref{eq:MJE}. 
However, this is not the case if the system is stochastic (the RHS would then be replaced by its average). Then in order to obtain Eq. \eqref{eq:proof}, one must use  the concept of time-reversed phase space trajectories, and invoke the Crooks relation for the ratio of forward to reverse path probabilities \cite{cro98_jsp}. 

{An application of Jensen's inequality to Eq. \eqref{eq:proof} yields the modified second law for feedback-driven systems:
  \begin{align}
    \av{W-\Delta F}\ge -k_BT\av{I}.
    \label{eq:secondlaw}
  \end{align}
We note that $\av{I}$ is a Kullback-Leibler divergence \cite{hor10_pre}, so the RHS of the above relation is negative.
Thus, it is in principle possible to beat the traditional second law (in which the right hand side of \eqref{eq:secondlaw} is zero) in feedback-driven systems. This happens, for instance, in the Szilard engine where work can be extracted in a cyclic process (by using thermal energy of the heat bath) even though there is no change in free energy. Even in the absence of feedback, the second law is valid on average, but transient violations are possible  at the level of a single trajectory \cite{eva02_prl,lah11_jpa}.

Let us consider the case when the initial condition is not thermal equilibrium, but is given by an arbitrary distribution $p(\bm x_0)$. We define the Kullback-Leibler distance between the actual distribution (given by the dynamics) and the corresponding equilibrium distribution (to which the system will relax if the external parameter is frozen in time) to be 
\begin{align}
  D(x,t)\equiv\int \bm{dx}~ p(\bm x,t)\ln\frac{p(\bm x,t)}{p^{eq}(\bm x,t)}.
\end{align}
Then the change in $D(x,t)$ for a trajectory is given by $\Delta D\equiv D(\bm x_\tau,\tau)-D(\bm x_0,0)$ \cite{lah15_injp}. 
 Given that the system would finally relax to an equilibrium state if the protocol is held fixed, one can now readily show that the Jarzynski equality generalizes to
\begin{align}
\av{e^{-\beta(W-\Delta F)+\Delta D}}=1.
\end{align}
 Application of Jensen's inequality gives
\begin{align}
  \av{W-\Delta F}\ge \Delta D.
\end{align}
This shows that even for a single temperature bath, one can extract work, provided the system is prepared initially in a non-equilibrium state.
}

We next show that the correction term (the mutual information in the case discussed above) is not unique, even if we do not use the concept of reverse trajectories. 

\section{Non-uniqueness of the correction term}

The other correction term that has been obtained in \cite{kun12_pre} is the following:
\begin{align}
I'=\ln\bigg[\frac{p(\bm y_0|\bm x_0)p(\bm y_1|\bm x_1)\cdots p(\bm y_N|\bm x_N)}{p(\bm y_0|\bm x_1)p(\bm y_1|\bm x_2)\cdots p(\bm y_N|\bm x_\tau)}\bigg].
\end{align}
To prove this, we proceed as before:
\begin{align}	
&\av{e^{-\beta(W-\Delta F)-I'}} \nn\\
 &= \int\{\bm{dx_i}\}\{\bm{dy_i}\}\frac{e^{-\beta H(0)}}{Z(0)}P_{\{\bm y_i\}}(\{\bm x_i\}|\bm x_0)\nn\\
 &~~~\times p(\bm y_0|\bm x_0)\cdots p(\bm y_N|\bm x_N)~e^{-\beta(H(\tau)-H(0))}\frac{Z(0)}{Z(\tau)}\nn\\
 &~~~\times \frac{p(\bm y_0|\bm x_1)\cdots p(\bm y_N|\bm x_\tau)}{p(\bm y_0|\bm x_0)\cdots p(\bm y_N|\bm x_N)}\nn\\
 &= \int \bm{dx}_0\{\bm{dy}_i\}\frac{e^{-\beta H(\tau)}}{Z(\tau)}p(\bm y_0|\bm x_1)\cdots p(\bm y_N|\bm x_\tau)\nn\\
 &= 1.
\end{align}
Here, we have used the normalization conditions $\int \bm{dy}_i p(\bm y_i|\bm x_{i+1})=1.$ 

We thus find that there can be more than one correction terms in the exponent. In fact, one can take a combination of $I$ and $I'$ in different parts of the trajectory and still get the answer as unity \cite{lah13_phyA}, which means that the number of possible corrections terms can be very large. At present it is not clear which one among them provides a better bound for the extracted work.

\section{The efficacy parameter}

The efficacy paremeter is defined as
\begin{align}
\gamma &\equiv \av{e^{-\beta(W-\Delta F)}}.
\end{align}
Note that $\gamma=1$ if the Jarzynski equality (Eq. \eqref{eq:JE}) is satisified (i.e. in absence of feedback). However, if the protocol is feedback-controlled the RHS is in general different from unity. We call this the efficacy parameter, because if the work dissipated $W_d \equiv W-\Delta F$ by the external agent is smaller, then it is larger. In this manner, $\gamma$ provides a measure of how efficient our feedback control is. For our system, we get
\begin{align}
\gamma &= \int \{\bm{dx}_i\}\{ \bm{dy}_i\} \frac{e^{-\beta H(0)}}{Z(0)}P_{\{\bm y_i\}}(\{\bm x_i\}|\bm x_0)\nn\\ 
& \hspace{0.5cm}\times p(\bm y_0|\bm x_0)p(\bm y_1|\bm x_1)\cdots p(\bm y_N|\bm x_N) \nn\\
& \hspace{0.5cm}\times e^{-\beta (H(\tau)-H(0))}\frac{Z(0)}{Z(\tau)} \nn\\
&= \int \{\bm{dx}_i\}\{ \bm{dy}_i\} \frac{e^{-\beta H^*(\tau)}}{Z(\tau)}P_{\{\bm y^*_i\}}(\{\bm x^*_i\}|\bm x^*_\tau)\nn\\
&\hspace{0.5cm} \times  p(\bm y_0^*|\bm x_0^*)p(\bm y_1^*|\bm x_1^*)\cdots p(\bm y_N^*|\bm x_N^*),
\end{align}
where asterisk (*) symbol implies time reversal operation: $x^*\equiv (\bm q,\bm p)^*=(\bm q,-\bm p)$. We have used the result $P_{\{\bm y_i\}}(\{\bm x_i\}|\bm x_0)=P_{\{\bm y^*_i\}}(\{\bm x^*_i\}|\bm x^*_\tau)$, {which means that the probability of tracing out a path $\{\bm x_i\}$ in phase space under protocol $\{\lambda_{\hat y_i}(t)\}$ for given initial point $\bm x_0$, is the same as that of tracing out a path $\{\bm x^*_i\}$ for a given initial point $\bm x^*_\tau$ and under the time-reversed protocol $\{\lambda_{\hat y_i}(\tau-t)\}$}. We have further used the fact that the error probabilities are invariant under time-reversal: $p(\bm y_i|\bm x_i) = p(\bm y^*_i|\bm x^*_i)$, and the symmetry of Hamiltonian: $H(\tau)=H^*(\tau)$. {Note that each $\bm x_i^*$ can be mapped to $\bm x_\tau^*$ through the relation $\bm x_i^*=M_{\tau-t_i}^{-1}(\bm x_\tau^*)$.} Using the compact notation 
\begin{align}
P^*(\{\bm x_i^*\},\{\bm y_i^*\})\equiv &
 \frac{e^{-\beta H^*(\tau)}}{Z(\tau)}P_{\{\bm y^*_i\}}(\{\bm x^*_i\}|\bm x^*_\tau)\nn\\
 &\times p(\bm y_0^*|\bm x_0^*)p(\bm y_1^*|\bm x_1^*)\cdots p(\bm y_N^*|\bm x_N^*),
\end{align}
one then arrives at the relation
\begin{align}
\gamma = \int \{ \bm{dy}_i\} P^*(\{\bm y_i\}).
\end{align}
The final expression is the total probability of obtaining time-reversed outcomes if measurements are performed on the time-reversed states. The efficacy parameter retains the same physical meaning irrespective of how the reverse process is generated.

\section{Extension to total entropy}

Till now we have focussed on systems that begin in a state of thermal equilibrium. 
For a system beginning from an arbitrary initial distribution $p(\bm x_0)$ and ending in the final distribution $p(\bm x_\tau)$, the change in system entropy is defined as $\Delta s=\ln[p(\bm x_0)/p(\bm x^*_\tau)]$ \cite{sei05_prl,sei08_epjb,sei12_rpp}. For a system undergoing Hamiltonian evolution, this itself is the total entropy produced during the process. This is because the system is an isolated one, and it cannot dissipate any heat into its surroundings. One can then check that
\begin{align}
  \av{e^{-\Delta s-I}} 
  &= \int \{\bm{dx}_i\}\{ \bm{dy}_i\} p(\bm x_0)P_{\{\bm y_i\}}(\{\bm x_i\}|\bm x_0)\nn\\ 
  & \hspace{0.5cm}\times p(\bm y_0|\bm x_0)p(\bm y_1|\bm x_1)\cdots p(\bm y_N|\bm x_N) \frac{p(\bm x^*_\tau)}{p(\bm x_0)}\nn\\
  &\hspace{0.5cm}\times \frac{P(\{\bm y_i\})}{p(\bm y_0|\bm x_0)p(\bm y_1|\bm x_1)\cdots p(\bm y_N|\bm x_N)}\nn\\
  &= \int \{\bm{dx}_i\}\{ \bm{dy}_i\} p(\bm x^*_\tau)P_{\{\bm y^*_i\}}(\{\bm x^*_i\}|\bm x_\tau^*)P(\{\bm y_i\})\nn\\
  &=1.
\end{align}
Once again, application of Jensen's inequality gives $\av{\Delta s}\ge -\av{I}$, which is the modified second law. 

We next derive the expression for the efficacy parameter in this case, which  is defined as
\begin{align}
  \gamma &= \av{e^{-\Delta s}} \nn\\
  &=\int \{\bm{dx}_i\}\{ \bm{dy}_i\}p(\bm x_0)P_{\{\bm y_i\}}(\{\bm x_i\}|\bm x_0)\frac{p(\bm x_\tau^*)}{p(\bm x_0)} \nn\\
  &\hspace{0.5cm}\times p(\bm y_0|\bm x_0)p(\bm y_1|\bm x_1)\cdots p(\bm y_N|\bm x_N)\nn\\
  &= \int \{\bm{dx}_i\}\{ \bm{dy}_i\}p(\bm x_\tau^*)P_{\{\bm y^*_i\}}(\{\bm x^*_i\}|\bm x_\tau^*)\nn\\
  &\hspace{0.5cm}\times p(\bm y_0^*|\bm x_0^*)p(\bm y_1^*|\bm x_1^*)\cdots p(\bm y_N^*|\bm x_N^*) \nn\\
  &= \int \{\bm{dx}_i\}\{ \bm{dy}_i\} P^*(\{\bm x_i^*\},\{\bm y_i^*\})\nn\\
  &= \int \{ \bm{dy}_i\} P^*(\{\bm y_i\}).
\end{align}
{Thus, we arrive at the same expression for the efficacy parameter that carries the same physical meaning, although our initial distribution is now considered to be arbitrary.}

\section{Extension to quantum systems}

For isolated quantum systems that undergo intermediate projective measurements, the above formalism can easily be generalized, since the evolution is unitary in-between any two measurements \cite{lah12_jpa,lah12_pramana,lah13_pramana}. There is one-to-one correspondence between the initial and the final quantum states during this unitary evolution, which is a necessary requirement for the derivations carried out above. The derivations are similar to the ones for classical Hamiltonian evolution, and therefore are not reproduced here. Our final results remain the same.

\section{Conclusions}

We have shown that the extended fluctuation theorems for work and total entropy in presence of multiple measurements and feedback can be derived for a Hamiltonian system, by considering the forward phase space trajectories only. The derivations are based on the fact that once the initial point is specified, the full trajectory gets specified for such systems. On the contrary, in stochastic systems, we must necessarily use the concept of reverse trajectories. We find that the correction term that appears in the exponent is not unique, but can be an arbitrary combination of $I$ and $I'$, as elaborated in the text. To derive the final expression for the efficacy parameter, however, the notion of reverse trajectory must be considered. We find that this expression retains the same physical meaning, irrespective of how the reverse trajectory is generated.

\section{Acknowledgement}

One of us (AMJ) thanks DST, India for financial support. SL is supported by the Region \^Ile-de-France thanks to the ISC-PIF.

\bibliographystyle{apsrev4-1}
\bibliography{ref}
\end{document}